
%


\def\oneandathirdspace{\baselineskip=\normalbaselineskip
  \multiply\baselineskip by 4 \divide\baselineskip by 3}

\def\doublespace{\baselineskip=\normalbaselineskip \multiply\baselineskip by 2}
\parskip=\medskipamount
\overfullrule=0pt
\raggedbottom
\def\normalparindent{24pt}
\nopagenumbers
\footline={\ifnum\pageno=1{\hfil}\else{\hfil\rm\folio\hfil}\fi}
\def\endpage{\vfill\eject}
\def\beginlinemode{\endmode\begingroup\parskip=0pt
                   \obeylines\def\\{\par}\def\endmode{\par\endgroup}}
\def\beginparmode{\endmode\begingroup \def\endmode{\par\endgroup}}
\let\endmode=\par
\def\raggedcenter{
                  \leftskip=2em plus 6em \rightskip=\leftskip
                  \parindent=0pt \parfillskip=0pt \spaceskip=.3333em
                  \xspaceskip=.5em\pretolerance=9999 \tolerance=9999
                  \hyphenpenalty=9999 \exhyphenpenalty=9999 }
\def\\{\cr}
\let\rawfootnote=\footnote\def\footnote#1#2{{\parindent=0pt\parskip=0pt
        \rawfootnote{#1}{#2\hfill\vrule height 0pt depth 6pt width 0pt}}}
\def\title{\null\vskip 3pt plus 0.2fill\beginlinemode\raggedcenter\bf}
\def\author{\vskip 3pt plus 0.2fill \beginlinemode\raggedcenter}
\def\affil{\vskip 3pt plus 0.1fill\beginlinemode\raggedcenter\it}
\def\abstract{\vskip 3pt plus 0.3fill \beginparmode  }
\let\endtopmatter=\endtitlepage
\def\body{\beginparmode\parindent=\normalparindent}
\def\head#1{\par\goodbreak{\immediate\write16{#1}
           {\noindent\bf #1}\par}\nobreak\nobreak}
\def\subhead#1{\par\goodbreak{\immediate\write16{#1}
           {\noindent\bf #1}\par}\nobreak\nobreak}
\def\refto#1{$^{[#1]}$}
\def\ref#1{Ref.~#1}
\def\Ref#1{Ref.~#1}\def\cite#1{{#1}}\def\[#1]{[\cite{#1}]}

\def\(#1){(\call{#1})}
\def\call#1{{#1}}\def\taghead#1{{#1}}
\def\references{\head{REFERENCES}\beginparmode\frenchspacing\parskip=0pt}
\gdef\refis#1{\item{#1.\ }}
\gdef\journal#1,#2,#3,#4.{{ #1~}{\bf #2}, #3 (#4)}
\def\endreferences{\body}
\def\endit{\endmode\vfill\supereject}\let\endpaper=\endit

\def\prd{\journal Phys. Rev. D,}
\def\prl{\journal Phys. Rev. Lett.,}

\def\npb{\journal Nucl. Phys. B,}

\def\plb{\journal Phys. Lett. B,}

\def\cqg{\journal Class. Quantum Grav.,}

\def\gsim{\mathrel{\raise.3ex\hbox{$>$\kern-.75em\lower1ex\hbox{$\sim$}}}}
\def\lsim{\mathrel{\raise.3ex\hbox{$<$\kern-.75em\lower1ex\hbox{$\sim$}}}}
\def\square{\kern1pt\vbox{\hrule height 0.6pt\hbox{\vrule width 0.6pt\hskip 3pt
   \vbox{\vskip 6pt}\hskip 3pt\vrule width 0.6pt}\hrule height 0.6pt}\kern1pt}
\def\sla{\raise.15ex\hbox{$/$}\kern-.72em}

\catcode`@=11
\newcount\r@fcount \r@fcount=0\newcount\r@fcurr
\immediate\newwrite\reffile\newif\ifr@ffile\r@ffilefalse
\def\w@rnwrite#1{\ifr@ffile\immediate\write\reffile{#1}\fi\message{#1}}
\def\writer@f#1>>{}
\def\referencefile{\r@ffiletrue\immediate\openout\reffile=\jobname.ref%
  \def\writer@f##1>>{\ifr@ffile\immediate\write\reffile%
    {\noexpand\refis{##1} = \csname r@fnum##1\endcsname = %
     \expandafter\expandafter\expandafter\strip@t\expandafter%
     \meaning\csname r@ftext\csname r@fnum##1\endcsname\endcsname}\fi}%
  \def\strip@t##1>>{}}

\def\citeall#1{\xdef#1##1{#1{\noexpand\cite{##1}}}}
\def\cite#1{\each@rg\citer@nge{#1}}
\def\each@rg#1#2{{\let\thecsname=#1\expandafter\first@rg#2,\end,}}
\def\first@rg#1,{\thecsname{#1}\apply@rg}
\def\apply@rg#1,{\ifx\end#1\let\next=\relax%
\else,\thecsname{#1}\let\next=\apply@rg\fi\next}%
\def\citer@nge#1{\citedor@nge#1-\end-}
\def\citer@ngeat#1\end-{#1}
\def\citedor@nge#1-#2-{\ifx\end#2\r@featspace#1
  \else\citel@@p{#1}{#2}\citer@ngeat\fi}
\def\citel@@p#1#2{\ifnum#1>#2{\errmessage{Reference range #1-#2\space is bad.}
    \errhelp{If you cite a series of references by the notation M-N, then M and
    N must be integers, and N must be greater than or equal to M.}}\else%
{\count0=#1\count1=#2\advance\count1 by1\relax\expandafter\r@fcite\the\count0,%
  \loop\advance\count0 by1\relax
    \ifnum\count0<\count1,\expandafter\r@fcite\the\count0,%
  \repeat}\fi}
\def\r@featspace#1#2 {\r@fcite#1#2,}    \def\r@fcite#1,{\ifuncit@d{#1}
    \expandafter\gdef\csname r@ftext\number\r@fcount\endcsname%
    {\message{Reference #1 to be supplied.}\writer@f#1>>#1 to be supplied.\par
     }\fi\csname r@fnum#1\endcsname}
\def\ifuncit@d#1{\expandafter\ifx\csname r@fnum#1\endcsname\relax%
\global\advance\r@fcount by1%
\expandafter\xdef\csname r@fnum#1\endcsname{\number\r@fcount}}
\let\r@fis=\refis   \def\refis#1#2#3\par{\ifuncit@d{#1}%
    \w@rnwrite{Reference #1=\number\r@fcount\space is not cited up to now.}\fi%
  \expandafter\gdef\csname r@ftext\csname r@fnum#1\endcsname\endcsname%
  {\writer@f#1>>#2#3\par}}
\def\r@ferr{\endreferences\errmessage{I was expecting to see
\noexpand\endreferences before now;  I have inserted it here.}}
\let\r@ferences=\references
\def\references{\r@ferences\def\endmode{\r@ferr\par\endgroup}}
\let\endr@ferences=\endreferences
\def\endreferences{\r@fcurr=0{\loop\ifnum\r@fcurr<\r@fcount
    \advance\r@fcurr by 1\relax\expandafter\r@fis\expandafter{\number\r@fcurr}%
    \csname r@ftext\number\r@fcurr\endcsname%
  \repeat}\gdef\r@ferr{}\endr@ferences}
\let\r@fend=\endpaper\gdef\endpaper{\ifr@ffile
\immediate\write16{Cross References written on []\jobname.REF.}\fi\r@fend}
\catcode`@=12
\citeall\refto\citeall\ref\citeall\Ref
\catcode`@=11
\newcount\tagnumber\tagnumber=0
\immediate\newwrite\eqnfile\newif\if@qnfile\@qnfilefalse
\def\write@qn#1{}\def\writenew@qn#1{}
\def\w@rnwrite#1{\write@qn{#1}\message{#1}}
\def\@rrwrite#1{\write@qn{#1}\errmessage{#1}}
\def\taghead#1{\gdef\t@ghead{#1}\global\tagnumber=0}
\def\t@ghead{}\expandafter\def\csname @qnnum-3\endcsname
  {{\t@ghead\advance\tagnumber by -3\relax\number\tagnumber}}
\expandafter\def\csname @qnnum-2\endcsname
  {{\t@ghead\advance\tagnumber by -2\relax\number\tagnumber}}
\expandafter\def\csname @qnnum-1\endcsname
  {{\t@ghead\advance\tagnumber by -1\relax\number\tagnumber}}
\expandafter\def\csname @qnnum0\endcsname
  {\t@ghead\number\tagnumber}
\expandafter\def\csname @qnnum+1\endcsname
  {{\t@ghead\advance\tagnumber by 1\relax\number\tagnumber}}
\expandafter\def\csname @qnnum+2\endcsname
  {{\t@ghead\advance\tagnumber by 2\relax\number\tagnumber}}
\expandafter\def\csname @qnnum+3\endcsname
  {{\t@ghead\advance\tagnumber by 3\relax\number\tagnumber}}
\def\equationfile{\@qnfiletrue\immediate\openout\eqnfile=\jobname.eqn%
  \def\write@qn##1{\if@qnfile\immediate\write\eqnfile{##1}\fi}
  \def\writenew@qn##1{\if@qnfile\immediate\write\eqnfile
    {\noexpand\tag{##1} = (\t@ghead\number\tagnumber)}\fi}}
\def\callall#1{\xdef#1##1{#1{\noexpand\call{##1}}}}
\def\call#1{\each@rg\callr@nge{#1}}
\def\each@rg#1#2{{\let\thecsname=#1\expandafter\first@rg#2,\end,}}
\def\first@rg#1,{\thecsname{#1}\apply@rg}
\def\apply@rg#1,{\ifx\end#1\let\next=\relax%
\else,\thecsname{#1}\let\next=\apply@rg\fi\next}
\def\callr@nge#1{\calldor@nge#1-\end-}\def\callr@ngeat#1\end-{#1}
\def\calldor@nge#1-#2-{\ifx\end#2\@qneatspace#1 %
  \else\calll@@p{#1}{#2}\callr@ngeat\fi}
\def\calll@@p#1#2{\ifnum#1>#2{\@rrwrite{Equation range #1-#2\space is bad.}
\errhelp{If you call a series of equations by the notation M-N, then M and
N must be integers, and N must be greater than or equal to M.}}\else%
{\count0=#1\count1=#2\advance\count1 by1\relax\expandafter\@qncall\the\count0,%
  \loop\advance\count0 by1\relax%
    \ifnum\count0<\count1,\expandafter\@qncall\the\count0,  \repeat}\fi}
\def\@qneatspace#1#2 {\@qncall#1#2,}
\def\@qncall#1,{\ifunc@lled{#1}{\def\next{#1}\ifx\next\empty\else
  \w@rnwrite{Equation number \noexpand\(>>#1<<) has not been defined yet.}
  >>#1<<\fi}\else\csname @qnnum#1\endcsname\fi}
\let\eqnono=\eqno\def\eqno(#1){\tag#1}\def\tag#1$${\eqnono(\displayt@g#1 )$$}
\def\aligntag#1\endaligntag  $${\gdef\tag##1\\{&(##1 )\cr}\eqalignno{#1\\}$$
  \gdef\tag##1$${\eqnono(\displayt@g##1 )$$}}
\def\eqalignno#1{\displ@y \tabskip\centering
  \halign to\displaywidth{\hfil$\displaystyle{##}$\tabskip\z@skip
    &$\displaystyle{{}##}$\hfil\tabskip\centering
    &\llap{$\displayt@gpar##$}\tabskip\z@skip\crcr
    #1\crcr}}
\def\displayt@gpar(#1){(\displayt@g#1 )}
\def\displayt@g#1 {\rm\ifunc@lled{#1}\global\advance\tagnumber by1
        {\def\next{#1}\ifx\next\empty\else\expandafter
        \xdef\csname @qnnum#1\endcsname{\t@ghead\number\tagnumber}\fi}%
  \writenew@qn{#1}\t@ghead\number\tagnumber\else
        {\edef\next{\t@ghead\number\tagnumber}%
        \expandafter\ifx\csname @qnnum#1\endcsname\next\else
        \w@rnwrite{Equation \noexpand\tag{#1} is a duplicate number.}\fi}%
  \csname @qnnum#1\endcsname\fi}
\def\ifunc@lled#1{\expandafter\ifx\csname @qnnum#1\endcsname\relax}
\let\@qnend=\end\gdef\end{\if@qnfile
\immediate\write16{Equation numbers written on []\jobname.EQN.}\fi\@qnend}
\catcode`@=12
\magnification=1200
\oneandathirdspace

\title
Quantum electrodynamics in the gravitational field of a cosmic string
\author
Vladimir D. Skarzhinsky$^{1,3}$, Diego D. Harari$^2$ and Ulf Jasper$^3$
\affil $^1$P. N. Lebedev Physical Institute
Leninsky prospect 53, Moscow 117924, Russia
\affil
$^2$Departamento de F\'\i sica, Facultad de Ciencias Exactas y Naturales
Universidad de Buenos Aires, Ciudad Universitaria - Pab.1
1428 Buenos Aires, Argentina
\affil $^3$Fakult\"at f\"ur Physik der Universit\"at Konstanz
Postfach 5560, D 78434 Konstanz, Germany
\abstract
We evaluate the cross section for electron-positron pair
production by a single high energy photon in the space-time of a
static, straight cosmic string. Energy and momentum conservation
precludes this process in empty Minkowski space. It happens around a
cosmic string, in spite of the local flatness of the metric, as a
consequence of the conical structure of space. Previous results based
on a simplified model with scalar fields are here extended to the
realistic QED case. Analytic expressions are found in three different
regimes: near the threshold, at energies much larger than the
electron
rest mass M, and at ultrahigh energies, much larger than $M/\delta$,
with $\delta$ the string mass per unit length in Planck units.

\affil To appear in Physical Review D

\endtopmatter

\head{1. Introduction}

Electron-positron pair production by a single photon with energy
higher than twice the electron rest mass does not occur in empty
Minkowski space, as dictated by conservation of linear momentum. It
is a rather common process, though, in the vicinity of atomic nuclei,
their inhomogeneous Coulomb field providing the requisite momentum.
The cross section for this process is known since the earlier times
of QED.\refto{Bethe34} The aim of this paper is to evaluate the cross
section for pair production in the gravitational field of a cosmic
string, and to discuss some of its potential consequences.

There is no Newtonian gravitational field around a static, straight
cosmic string. The metric is locally flat. One might naively suspect
impossible to have pair production by a single photon. This, however,
is not the case, as was shown in a previous paper\refto{HS90} in the
context of a simplified model based on a scalar field theory. The
metric around a static straight string that lies along the z-axis
reads, in cylindrical coordinates:\refto{Vilenkin85, Gott85}
$$ds^2=dt^2-d\rho^2-\rho^2d\theta^2-dz^2\;.\eqno(m)$$
The metric is the same as in Minkowski space, but here the
periodicity
of the angular coordinate is within the range
$$0\leq\theta\leq
{2\pi\over\nu},\quad{\rm with}\quad\nu= (1-4G\mu)^{-1}.\eqno(d)$$
$\mu$ is the mass per unit length of string, and G is Newton's
constant. The dimensionless quantity G$\mu$ measures the strength of
the gravitational effects of the string. The space-like sections
around the string have the topology of a cone with the vertex at the
core and with deficit angle $8\pi$G$\mu$. The conical structure of
the cosmic string space-time is the source of momentum
non-conservation
in the plane perpendicular to the string, which allows for pair
production by a single photon. This absence of global momentum
conservation was already stressed for gravity in 2+1
dimensions,\refto{Henneaux84} where massive particles originate a
conical structure of space-time,\refto{Deser84} the same as strings
do in 3+1 dimensions.

The gravitational mechanism that permits pair production by a single
photon around a cosmic string bears some resemblance with the
Aharonov-Bohm effect,\refto{Aharonov59} at least in its topological
features. Recently the Aharonov-Bohm interaction of fermions with the
 pure gauge potentials around  cosmic strings has been shown to lead
to significantly large scattering cross sections in models where the
strings carry non-integer
fluxes.\refto{Alford89a,Alford89b,Gerbert89b,Perkins91b}
The gravitational effect, though, is of a different nature than the
Aharonov-Bohm effect, and is independent on the details of the field
theoretical model for the cosmic string.

Classically,
the gravitational effects of a straight cosmic string
become manifest when two particles move along opposite sides of the
string. Initially parallel trajectories converge as they move past
the string, which then acts as a gravitational
lens.\refto{Vilenkin81,Vilenkin85} The deflection angle is
independent
on the impact parameter. This peculiar features lead to the picture
of a classical analogue of the Aharonov-Bohm effect.\refto{Ford81}
Other classical effects may affect a
single  particle: the conical structure of the space-time induces
a self-force, both gravitational as well as electrostatic, on a test
particle around a string.\refto{Linet86,Smith90} Also, a freely
moving
charged particle radiates as it goes by a string,\refto{Frolov88} in
analogy to the radiation by a charged particle when it suffers
Aharonov-Bohm scattering.\refto{Serebryany89}
It must be noted that the conical singularity of the cosmic string
metric \(m), \(d) is smoothed out in more realistic string models. In
this sense the effects under consideration are not of topological
origin but are  caused by matter fields concentrated in cosmic strings.

This classical gravitational effects around a cosmic string give us a
 hint about the mechanism underlying pair production by a single
photon. An heuristic, semiclassical picture follows. Virtual
positron-electron
pairs are continuously created from the vacuum, and quickly
annihilate. A photon is not able to make them real in empty Minkowski
space, even if it has sufficient energy, since otherwise momentum
conservation would be violated. But in the presence of the string,
the
requisite momentum to make them real is provided by the string if
each
member of the pair moves along opposite sides. This picture may prove
helpful to qualitatively understand some of the quantitative results
we shall derive. Further properties of the pair production process,
as well as extensions to others such as bremsstrahlung, were
considered
in Ref.\cite{Audretsch91} in the context of a scalar field theory.
These problems in the framework of QED will be treated in a
separate paper.\refto{Audretsch93}

This paper is organized as follows. In sections 2 and 3 we set up the
stage by discussing the quantization of free electron and photon
fields respectively in a cosmic string space-time. In section 4 we
write up the QED interaction Lagrangian in the cosmic string
gravitational background, and derive a closed expression for the
matrix elements and cross section for pair production by a single
high
energy photon. We present analytic approximations valid at different
energy regimes. In section 5 we round up the conclusions.

\head{2. Dirac fields in a cosmic string space-time}

In this section we find the appropriate field operator for electrons
and positrons in the conical space around a straight comic string.
Similar studies of the Dirac equation were performed in 2+1
dimensions, with two-component spinors, both in a conical
space\refto{Gerbert89a} as well as in a vortex background, relevant
for the Aharonov-Bohm
effect.\refto{Alford89a,Alford89b,Gerbert89b,Perkins91b,Hagen90}
We shall work
instead with usual four-component spinors in the 3+1 dimensional
space-time around a cosmic string, closer to the treatment of the
Aharonov-Bohm effect performed in Ref.\cite{Voropaev91}.

Dirac equation in a curved space-time reads:\refto{Weinberg72}
$$
(i \gamma^{\mu}(x){\widetilde\nabla_{\mu}} - M)\psi = 0\;,\eqno(de)$$
with $\gamma^{\mu}(x) = V^{\mu}_{(a)}\gamma^{a},$
where $V^{\mu}_{(a)}$ is a tetrad, and $\gamma^a$ are ordinary
Dirac matrices, that satisfy
$\gamma^a\gamma^b + \gamma^b\gamma^a = 2\eta^{ab}.$
We work with the representation
$$\gamma^{0} = \pmatrix {1 & 0 \cr
                       0 & -1 \cr}, \;
                       \gamma^{i} =  \pmatrix {0 & \sigma^{i} \cr
                                              -\sigma^{i} & 0
\cr}\;.\eqno(gammas)$$
In eq.\(de) the covariant derivative involves the spin
connection $\Gamma_\mu$
$$
\widetilde\nabla_{\mu} = \partial_{\mu} + \Gamma_{\mu}\;,\quad
\Gamma_{\mu} = {1 \over2}\Sigma^{ab}V^{\nu}_{(a)}(\nabla_{\mu}V_{(b)
\nu})\;,\quad \Sigma^{ab} = {1
\over4}[\gamma^{a},\gamma^{b}]\;.
\eqno(covder)$$
For the cosmic string metric \(m) we can take the tetrad as
$$
V^{\mu}_{(a)} = \pmatrix { 1 &  0  &  0  &  0 \cr
                           0 &  c  &  s  &  0 \cr
                           0 & -{s/\rho} &{ c/\rho} & 0 \cr
                           0 &  0  &  0  &  1 \cr}\eqno(tetrad)$$
where $c\equiv\cos(\nu \theta), s\equiv\sin(\nu \theta).$
Then the Dirac matrices $\gamma^\mu(x)$ in cylindrical coordinates
read
$$
\gamma^{0}(x) = \gamma^{0}\;, \; \gamma^{3}(x) = \gamma^{3}\;, \;
\gamma^{\rho}(x) = c\gamma^{1} + s\gamma^{2}\;, \; \gamma^{\theta}(x)
 =
{1 \over\rho}(-s\gamma^{1} + c\gamma^{2})
\eqno(gammac)$$
and the spin connection is
$$
\Gamma_{\mu} = \delta^{\theta}_{\mu}{\nu - 1 \over
2}{\gamma^{2}\gamma^{1}} = \delta^{\theta}_{\mu}{{\nu - 1} \over 2}i
\Sigma_{3}\;.\eqno(spincon)$$
Finally, the Dirac equation in the cosmic string metric reads
$$
\left( i[\gamma^{0}\partial_{t} + \gamma^{3}\partial_{z} +
\gamma^{\rho}(\partial_{\rho} - {{\nu -1} \over {2\rho}}) +
\gamma^{\theta}\partial_{\theta}] - M \right) \psi = 0\;.
\eqno(dirac)$$
It is useful to write it also in the Hamiltonian form:
$$
i\partial_{t}\psi = H\psi\;,\quad{\rm with} \quad
H = \alpha^{i}p_{i} + \beta M \eqno(ham)$$
where
$$
\alpha^{i} = \pmatrix{      0    & \sigma^{i} \cr
                      \sigma^{i} &       0 \cr}\;, \quad
                      \beta = \gamma_{0}\;,$$
and
$$
p_{\rho} = - i\partial_{\rho}\;,\quad\quad p_{\theta} = L_{3}\equiv
-i\partial_{\theta} + {\nu - 1 \over 2}{ \Sigma_{3}}\;,\quad\quad
p_{3} = -i\partial_{z}\;.$$
Here $\alpha^\rho$ and $\alpha^\theta$ are defined from $\alpha^1$
and
$\alpha^2$ as for the Dirac matrices in \(gammac). We can now write
a complete set of commuting operators, and use their eigenvalues to
label the quantum states of electrons and positrons:
$$
\eqalign{
&\hat{H}\psi = E\psi\;,\cr
&\hat{p_3}\psi = p_3\psi\;,\cr
&\hat{J_3}\psi = j_3\psi\;,\quad
\hat{J_3} = L_3 + {1\over 2} \Sigma_3\;,\quad  j_3 = \nu j\;,\cr
&\hat{S_t}\psi = s\psi\;,\quad \hat{S_t} = {1\over\sqrt{E^2 -
M^2}}\;\Sigma_i p_i \;.\cr}\eqno(opd)$$
The solution to these equations representing an electron state with
positive energy is
$$
\psi_{e}(j,x) = {\sqrt{\nu} \over 2\pi}N_{e}\exp(-iEt +
ip_3z)\pmatrix{ u \cr
v\cr}\;,\eqno(el)$$
with the two-component spinors $u,v$ given by
$$
u = {1 \over \sqrt{E - M}}\cdot
\pmatrix{ J_{\beta_1}(p_{\perp}\rho)\exp(i\nu l\theta)\cr
{i\epsilon_{l}sp_{\perp} \over p +
sp_3}J_{\beta_2}(p_{\perp}\rho)\exp(i\nu(l +1)\theta)\cr}$$
$$
v = {1 \over \sqrt{E + M}}\cdot
\pmatrix{ sJ_{\beta_1}(p_{\perp}\rho)\exp(i\nu l\theta)\cr
{i\epsilon_{l}p_{\perp} \over p +
sp_3}J_{\beta_2}(p_{\perp}\rho)\exp(i\nu(l +1)\theta)\cr}$$
and
$$
N_{e} = {\sqrt{p(p + sp_3)} \over{2\sqrt{E}}}\;,$$
$$
\beta_{1,2} = |\nu j \mp{1/2}|\;,\quad j = l + 1/2\;,$$
$$
p_{\perp} = \sqrt{p^2 - p^2_3} = \sqrt{E^2 - M^2 - p^2_3}\;,$$
$$
\epsilon_l = {\rm sign}(l)=\pm 1,\; \pmatrix{ l\geq 0,\cr
                                l<0 \cr}$$
and $s = \pm 1$ is the value of $S_{t}$. $J_\beta$ are Bessel
functions. Positron states $\psi_{p}$ with quantum numbers \(opd)
can be obtained from electron states of negative energy through
charge conjugation with the matrix  $C = \alpha^{2}$ and by changing
$p_3 \rightarrow -p_3, j \rightarrow -j \;(l \rightarrow -l -
1).$  So we find:
$$
\psi^{c}_{p}(j,x) = C\bar
\psi_{p,transp} = {\sqrt{\nu} \over 2\pi}N^{\ast}_{p}\exp(iEt -
ip_3z)\pmatrix{ y \cr w\cr}\;,\eqno(pos)$$
where
$$
y = {1 \over \sqrt{E + M}}\cdot
\pmatrix{ J_{\beta_2}(p_{\perp}\rho)\exp(-i\nu (l +1)\theta)\cr
{-i\epsilon_{l}sp_{\perp} \over p -
sp_3}J_{\beta_1}(p_{\perp}\rho)\exp(-i\nu l\theta)\cr}$$
$$
w = {1 \over \sqrt{E - M}}\cdot
\pmatrix{ -sJ_{\beta_2}(p_{\perp}\rho)\exp(-i\nu (l + 1)\theta)\cr
{i\epsilon_{l}p_{\perp} \over p - sp_3}
J_{\beta_1}(p_{\perp}\rho)\exp(-i\nu l\theta)\cr}$$
with
$$
N_{p} = {\sqrt{p(p - sp_3)} \over{2\sqrt{E}}}\;.$$
The electron-positron
field operator is $$ \psi(x,t) = \int d\mu_j [\psi_{e}(j,x) a_j +
\psi^{c}_{p}(j,x) b^{\dagger}_{j}]$$
where $a_j$ and $b_j$ are the
annihilation operators for electrons and positrons with given quantum
 numbers. The normalization condition is
$$
\int dx
\psi^{\dagger}_{e,p}(j,x) \psi_{e,p}(j^{\prime},x) = \delta_{j, j^{
\prime}}={\delta_{s, s^{\prime}}} {\delta_{l,  l^{ \prime}}}
{ \delta(p_{3} - p_3^\prime)} {\delta(p_{ \perp} -  p_{ \perp}^{
\prime}) \over
\sqrt{p_{\perp} p_{ \perp}^{ \prime}}}\;.\eqno(normd)$$
We denote collectively by  $j$ or $j^\prime$  the
quantum numbers of a given state,
and integration is also collectively denoted as
$$
\int d \mu_{j}
= { \sum_{l= - \infty}^{ \infty}}{ \int_{- \infty}^{  \infty}
{dp_{3}}}{ \int_{0}^{ \infty} p_{ \perp} dp_{ \perp}}\;. $$

Before closing this section, it is worth pointing out that although
the solution shares many similarities with that for fermions around
a magnetic flux tube, there are also significant
differences. An important distinction is that the parameter $\nu$,
which
deviates from unity in the conical space, appears multiplicatively in
the order of the Bessel functions, while in the Aharonov-Bohm case
the flux modifies additively  the order of the Bessel functions.
In connection with this, problems related with the non-self
adjointness of the Dirac operator in the Aharonov-Bohm
case\refto{Gerbert89b,Voropaev91,Kay91} do not seem to affect the
present discussion.

\head{3. Maxwell fields in a cosmic string space-time}

Maxwell equations in the cosmic string space-time with metric \(m),
in the Lorentz gauge, $\nabla_{\mu}A^{\mu} = 0,$ take the form
$$\eqalign{
\square {A_{\rho}} - {2 \over \rho^3}\partial_{\theta}A_{\theta} - {1
\over \rho^2}A_{\rho} =& 0\cr
\square {A_{\theta}} - {2 \over \rho}\partial_{\rho}A_{\theta} + {2
\over \rho}\partial_{\theta}A_{\rho} =& 0\cr
\square {A_{z}} = \square {A_{t}} =& 0\;,\cr}\eqno(max1)$$
where
$$
\square = \Delta - \partial^{2}_{t t}\;,\quad \Delta = {1 \over \rho}
 \partial_{\rho}( \rho \partial_{\rho}) + {1 \over
\rho^2}\partial^{2}_{\theta \theta} + \partial^{2}_{z z}\;.$$
In order to decouple these equations, so as to identify the linear
independent components to be quantized, we introduce spin-weighted
components, defining\refto{Aliev89}
$$
A_{\xi} = {1 \over \sqrt{2}}(A_{\rho} +{i\xi\over \rho}
A_{\theta})\quad {\rm if}\quad \xi=\pm 1\quad ;\quad
A_\xi=A_z,A_t\quad {\rm if}\quad\xi=0\;.\eqno(axi)$$
The equations for the $A_\xi$ components decouple, and can be
collectively
written as
$$
\square_{\xi}{A_{\xi}} = 0\;, \quad {\rm with}\quad
\square_{\xi} = \Delta_{\xi} -\partial^{2}_{t t}\;,\eqno(maxdec)$$
where
$$
\Delta_{\xi} = {1 \over \rho} \partial_{\rho}( \rho \partial_{\rho})
- {1 \over \rho^2}L^{2}_3 + \partial^{2}_{z z}\;,\quad
L_3 = -i\partial_{\theta} + \xi \;.\eqno(lap)$$
The normal modes for the independent components are
$$
f_{\xi}(j,x) = {\sqrt \nu \over 2 \pi} \exp(i \nu m \theta + ik_3z)
J_{|\nu m + \xi|}(k_{\perp} \rho){1 \over \sqrt{2\omega}}\exp(-i
\omega t)\eqno(nm)$$
where $j$ denotes collectively the values of $(k_3,k_\perp,m)$.
$k_3$ is the eigenvalue of $\hat k_{z} = -i\partial_{z}$, and
$l_3=\nu m+\xi$ is the eigenvalue of $L_3$. The modes are normalized
according to
$$
\int dx f^{\ast}_{\xi}(j,x)(i \buildrel \leftrightarrow \over
\partial_t) f_{\xi}(j^{\prime},x) = \delta_{j, j^{
\prime}}\;.\eqno(normmax)$$
The total solution to the Maxwell equations is
$$
A_{\xi}(t,x) = \int d\mu_j (f_{\xi}(j,x)c_{\xi} +
f^{\ast}_{-\xi}(j,x)c^{\dagger}_{-\xi})\;,\eqno(a)$$
or, back to cylindrical cordinates coordinates,
$$\eqalign{
A_{\rho}(t,x) =& \int d\mu_j {1 \over \sqrt{2}}[(c_{+}f_{+} +
c_{-}f_{-}) + (c^{\ast}_{+}f^{\ast}_{+} +
c^{\ast}_{-}f^{\ast}_{-})]\;,\cr
A_{\theta}(t,x) =& \int d\mu_j {-i \rho \over \sqrt{2}}[(c_{+}f_{+} -
 c_{-}f_{-}) - (c^{\ast}_{+}f^{\ast}_{+} -
c^{\ast}_{-}f^{\ast}_{-})]\;,\cr
A_{z,t}(t,x) =& \int d\mu_j [c_{3,0}f_{0} + c^{\ast}_{3,0}f^{\ast
}_{0}]\;.\cr}\eqno(acyl)$$
The solution with a given value of $\xi$ not only has a definite
quantum number $l_3 =\nu m + \xi$, but is also an eigenfunction
of the spin z-component operator
$$
S_3 = \pmatrix { 0    & -i/\rho & 0 \cr
                i\rho &   0     & 0 \cr
                 0    &   0     & 0 \cr}\eqno(s3)$$
with quantum number $s_3 = - \xi.$
So, the photon wave functions
$$
A^{+}_{i} = {1 \over \sqrt{2}}\pmatrix{ f_{+} \cr
                                -i \rho f_{+} \cr
                                           0, \cr}, \;
A^{-}_{i} = {1 \over \sqrt{2}}\pmatrix{ f_{-} \cr
                                 i \rho f_{-} \cr
                                           0, \cr}, \;
A^{0}_{i} = \pmatrix{ 0 \cr
                      0 \cr
                    f_{0} \cr}\eqno(ai)$$
have the following quantum numbers
$$
j_3 = \nu m\;, \quad l_3 = (\nu m \pm 1,\nu m)\; ,\quad s_3 = (\mp
1,0)\;.\eqno(jls)$$
The coefficients $c_{\xi}(j)$ can be promoted to operators with the
commutation relations
$$
[c_{\xi}(j), c^{\dagger}_{\xi}(j^{\prime})] = \delta_{j, j^{ \prime}}
 \;.\eqno(cr)$$
These operators are annihilation and creation operators for a photon
with quantum numbers $k_{\perp}, k_3, j_3(m), l_3(m), s_3.$

As a final remark, let us find out operators $c_{\sigma,\pi}$ which
correspond to annihilation operators for physical, transverse
photons,
by explicitly working out the Lorentz gauge condition, which
translates into:
$$
{k_{\perp} \over \sqrt{2}}[\delta_{m0}(c_{+} + c_{-}) +
\epsilon_{m}(c_{+} - c_{-})] + i(k_3 c_3 - \omega c_0) =
0\;,\eqno(lg)$$
where, as previously defined, $\epsilon_m={\rm sign}(m)$.
With $\vec{k}$ perpendicular to the string direction this determines
annihilation operators for transverse photons as $c_{\sigma}={1\over
\sqrt{2}}(c_{+}+c_{-})$ and $c_{\pi}=c_3.$

{\head{4. Cross section for pair production by a high energy photon}
\subhead{4.1. Matrix elements and cross section}}

In this section we evaluate the matrix elements and cross section for
 electron-positron pair production by a single high energy photon
moving in the flat but conical space-time around a cosmic string.
This
was already done for a simplified model based on scalar
fields.\refto{HS90,HS91,Audretsch91} The extension to QED is not so
straightforward, as we have seen already in the previous sections.
Actually, a very relevant difference  appears in the energy
dependence
of the cross section at high energies, which is larger than in the
corresponding scalar analog, and this may have significant
consequences.

The QED interaction Lagrangian is
$$
L_{int} = -e\sqrt{-g}\bar \psi(x)A_{\mu}(x)
\gamma^{\mu}(x)\psi(x)\eqno(qed)$$
and we can write, in the notation of the previous sections:
$$
A_{\mu}(x) \gamma^{\mu}(x) =
 \sqrt{2}A_{+} \exp(i\nu \theta) \gamma^{+} + \sqrt{2}A_{-}
 \exp(-i\nu \theta) \gamma^{-} + A_{z}(x) \gamma^3\;,\eqno(ag)$$
where
$$
\gamma^{\pm} = {1 \over 2}( \gamma^1 \mp i\gamma^2) = \pmatrix{
              0    & \sigma_{\pm} \cr
         - \sigma_{\pm} &       0 \cr}, \;
          \sigma_{+} = \pmatrix{ 0 & 0 \cr
                                 1 & 0 \cr}, \;
          \sigma_{-} = \pmatrix{ 0 & 1 \cr
                                 0 & 0 \cr}\;.\eqno(gammapm)$$
The matrix element for pair production of an electron with quantum
numbers $j_p = (p_{\perp}, p_3,l,s)$ and a positron with quantum
numbers $j_q = (q_{\perp}, q_3,n,r)$ by a single photon with quantum
numbers $j_k = (k_{\perp}, k_3,m, \lambda)$, can be written for
physical
states with $\lambda=\sigma,\;\pi$ in terms of matrix elements with $\xi =
\pm 1, 3,$
$$
M_{\xi} = <j_p, j_q|S^{(1)}|j_k> = - e\sqrt{2} \int {d^4
x}\bar\psi_{e}(j_{p},x) \gamma ^{\xi}
\psi_{p}^{c}(j_{q},x)f_{\xi}(j_{k},x) \exp(i\xi \nu
\theta)\eqno(me)$$
that after a simple integration on $t,z$ and $\theta$ become
$$
M_{\xi} = -e\sqrt{\nu}{{\sqrt{pq}} \over{4\sqrt{\omega E_{p}E_{q}}}}
{\delta(\omega -E_{p}-E_{q}) \delta(k_3 - p_3 - q_3)
\delta_{m,\;l + n + 1}} m_{\xi}\eqno(me2)$$
where
$$\eqalign{
m_+ =& -i\epsilon_l \sqrt{(p-sp_3)(q-rq_3)} R
J(\alpha_{+},\beta_{+}),\cr
m_- =& -i\epsilon_n sr \sqrt{(p+sp_3)(q+rq_3)} R
J(\alpha_-,\beta_-),\cr
m_3 =& {1\over\sqrt{2}}[r \sqrt{(p-sp_3)(q+rq_3)} J(\alpha_+,\beta_-)
+ s \sqrt{(p+sp_3)(q-rq_3)} J(\alpha_-,\beta_+)]R\cr}$$
with
$$
R={1\over\sqrt{E_p+M}\sqrt{E_q+M}}-{sr\over\sqrt{E_p-M}\sqrt{E_q-M}}.
$$
Here we denote
$$
J(\alpha,\beta) = \int_{0}^{ \infty}{d\rho}\rho
J_{|\alpha|}(p_{\perp}\rho)
J_{|\beta|}(q_{\perp}\rho)J_{|\alpha+\beta|}(k_{\perp}\rho)\eqno(jpm)
$$
with
$$
\alpha_{\pm} = \nu(l+{1\over2})\pm{1\over2},\;\beta_{\pm} =
\nu(n+{1\over2})\pm{1\over2}\;.$$
For $k_{\perp}>p_{\perp}+q_{\perp}$ we obtain \refto{Gradshteyn80}
$$
J(\alpha,\beta) = \Theta(-\alpha \beta){{2\sin[\pi
\hbox{min}(|\alpha|,|\beta|)]}\over{\pi k_{\perp}^2
\cos(A+B)\cos(A-B)}} {\left({\sin{A} \over \cos{B}}
\right)}^{|\alpha|}{\left({\sin{B} \over \cos{A}}
\right)}^{|\beta|}\eqno(jpm2)$$
where
$$
p_{\perp} =
k_{\perp}\sin{A}\cos{B},\;q_{\perp}=k_{\perp}\sin{B}\cos{A}\;.$$
We notice in \(me2) that energy as well as linear momentum along the
string direction are of course conserved. The condition $m=l+n+1$ is
the conservation law for the total angular momentum projection along
the string direction
$$
j_3 = \nu m = \nu (j + j')\;,\quad j=l+{1\over2}\;,\quad
j'=n+{1\over2}\;.\eqno(j3)$$
Notice in \(jpm2) the step function
$\Theta(-\alpha \beta) =
\Theta(-jj')$  in the matrix element. The members of the pair
produced by a high energy photon must have opposite signs
for their total angular momentum projections.
This is also the case for scalar particles.\refto{HS90}
An heuristic explanation for this fact arises in the framework of the
semiclassical picture presented in section 1.
Opposite signs for $j$ and $j^\prime$ represent, in a classical
description, motion along opposite sides of the string, which is
necessary to extract linear momentum away from it, and make
real pairs out from virtual vacuum fluctuations. This, in some sense,
implies a localization of the pair production mechanism in the
neighborhood of the string core, as also discussed with a different
approach in Ref.\cite{Audretsch91}

{}From the matrix element \(me) we evaluate the differential
probability
per unit length of string and per unit time for the pair production
process:
$$
W_{\lambda} = {{\nu e^2 pq}\over{64\pi^2\omega
E_{p}E_{q}}}{\delta(\omega -E_{p}-E_{q}) \; \delta(k_3-p_3-q_3) \;
\delta_{m,\;l+n+1}} |m_{\lambda}|^2\;,\eqno(Wxi)$$
where
$$\eqalign{
|m_{\sigma,\pi}|^2 &= {8(E_p E_q-M^2-srpq)\over \pi^2 k^4_{\perp}
p^2q^2\cos^{2}(A+B) \cos^{2}(A-B)}\cdot \cr
& \cdot[s\sqrt{(p-sp_3)(q\mp rq_3)} d(\alpha_{+},\beta_{\pm}) \mp
r\sqrt{(p+sp_3)(q\pm rq_3)} d(\alpha_{-},\beta_{\mp})]^2, \cr}$$
$$
d(\alpha,\beta) = \Theta(-\alpha \beta){\sin[\pi
\hbox{min}(|\alpha|,|\beta|)]} a^{|\alpha|\over 2}b^{|\beta|\over
2}\;,$$
$$
a = {\sin^{2}{A} \over \cos^{2}{B}}, \; b = {\sin^{2}{B} \over
\cos^{2}{A}}\;.$$
Notice that the differential probability $W_{\xi}$ depends on the
electron and positron polarizations mostly by the factor
$(E_pE_q+M^2-srpq)$ (and only in that way for pairs moving in the
plane perpendicular to the string). Hence, pairs with opposite
polarizations are produced with larger probability.

We are specially interested in the total cross section for pair
production.
To evaluate it, we need to integrate over final states.
The sum over polarizations of the created particles is
straightforward, and gives the factor
$$
\sum_{s,r}(p \mp sp_3) \;(q \mp rq_3)(E_{p}E_{q}+M^2-srpq) =
2pq(k^{2}_{\perp} - p^{2}_{\perp} - q^{2}_{\perp})\;.\eqno(sumpol1)$$
The sum over  angular momentum quantum numbers is more difficult to
perform, but at the end we obtain
$$\eqalign{
\sum_{l,n}\delta_{m,\;l+n+1} d^2(\alpha_{\xi},\beta_{\xi'})& =
\bigg\{\Theta(m)\left[a^{\xi\over 2}b^{-{\xi'\over 2}}a^{\nu
m}+a^{-{\xi\over 2}}b^{\xi'\over 2}b^{\nu m}\right]\cr
&+\Theta(-m)\left[a^{-{\xi\over 2}}b^{\xi'\over 2}a^{-\nu
m}+a^{\xi\over 2}
b^{-{\xi'\over 2}} b^{-\nu m}\right]\bigg\}(ab)^{\nu \over 2}
\Sigma\cr}\eqno(sumpol2)$$
where
$$
\Sigma = \sum_{l=0}^{\infty} \cos^{2}[\pi \nu (l-{1\over2})]
(ab)^{\nu l} = \cos^{2}{{\pi \nu}\over 2}{{(1-a^{\nu}b^{\nu})^{2}
+ 8a^{\nu}b^{\nu}\sin^{2}{{\pi \nu}\over 2}} \over
{(1-a^{\nu}b^{\nu}) [(1-a^{\nu}b^{\nu})^{2} -
4a^{\nu}b^{\nu}\sin^{2}{\pi \nu]}}}\;.$$
Notice the factor $\cos^{2}{{\pi \nu}\over 2}$ that vanishes, as it
should, for $\nu=1$, when there is no deficit angle.
It is different than the analogous factor $\sin^{2}(\pi \nu)$
that appears in the case of pair production of scalar
particles,\refto{HS90} and of the Aharonov-Bohm case
(with the number of magnetic flux units playing the role of $\nu$).
This distiction may be understood as the influence of the spin
connection on the Dirac equation in curved metrics.

Next we integrate on $dp_{3}dq_{3}$
$$\eqalign{
\int_{-\infty}^{\infty}{dp_3}\int_{-\infty}^{\infty}{dq_3}&{1\over{E_
{p}E_{q}}} {\delta(\omega -E_{p}-E_{q}) \delta(k_3-p_3-q_3)} =\cr
&{4 \Theta \left [k^{4}_{\perp}-
2k^{2}_{\perp}(p^{2}_{\perp}+q^{2}_{\perp}) +
(p^{2}_{\perp}-q^{2}_{\perp})^{2} -4k^{2}_{\perp}M^{2} \right ]}
\over{\sqrt{ k^{4}_{\perp}- 2k^{2}_{\perp}(p^{2}_{\perp}+
q^{2}_{\perp}) +
(p^{2}_{\perp}-q^{2}_{\perp})^{2}-
4k^{2}_{\perp}M^{2}}}\;\cr}\eqno(intpq)$$
and then we average on photon polarizations,
$$
{1\over
2}\sum_{l,n}\delta_{m,\;l+n+1}\sum_{\xi,\xi'}d^2(\alpha_{\xi},\beta_{
\xi'})
={{a+b+1+ab}\over\sqrt{ab}}{(ab)^{\nu\over 2}}{{a^{\nu |m|}+b^{\nu
|m|}\over 2}}\Sigma\;.\eqno(av)$$
The final step would be the integration on
$p_{\perp}$ and $q_{\perp}$, which can not be done analytically
for arbitrary incoming photon energy. Thus we now introduce a
convenient parametrization that facilitates analytic
approximations in different energy regimes.
We define the variables
$$r={p_\perp^2+q_\perp^2\over k_\perp^2}\;,\quad
t={p_\perp^2-q_\perp^2\over
p_\perp^2+q_\perp^2}\;,\quad s=\sqrt{1-2r+r^2t^2}\;. \eqno(rts)$$
Integrated on $t$ and $s$, the expression for the total cross section

takes the form
$$
\sigma_m = {{2 \nu e^{2}}\over{\pi^{4}\omega}}{\cos^{2}{{\pi
\nu}\over2}}
\int_{0}^{1}{dt} \int_{\epsilon}^{1}{ds} B(s,t)
f_{\nu}(a,b)\eqno(sigma)$$
where $\epsilon \equiv 2M/k_{\perp}$ and
$$
B(s,t) = {1\over \sqrt{1-t^2}}{1\over s\sqrt{s^2-{\epsilon}^2}}
{s^2+\sqrt{1-t^2(1-s^2)}\over[1+\sqrt{1-t^2(1-s^2)}]^2}
\left[1-{s^2(1-\sqrt{1-t^2(1-s^2)})\over
2\sqrt{1-t^2(1-s^2)}} \right]$$
$$
f_{\nu}(a,b) = {{a^{ \nu |m|} + b^{ \nu |m|} \over 2}}{(ab)^{ \nu
\over 2}} {{(1-a^{\nu}b^{\nu})^{2} + 8a^{\nu}b^{\nu}\sin^{2}{{\pi
\nu}\over 2}} \over {(1-a^{\nu}b^{\nu}) [(1-a^{\nu}b^{\nu})^{2} -
4a^{\nu}b^{\nu}\sin^{2}{\pi \nu]}}}\;.$$
In terms of $r,t$ and $s$, the functions $a$ and $b$ read
$$a={1+rt-s\over 1+rt+s}\;,\quad
b= {1-rt-s\over 1-rt+s}\;. \eqno(ab)$$
Eq. \(sigma) is our final closed expression for the cross section.

\subhead{4.2. Approximations at different energy regimes}

We now analyze the expression \(sigma) under different
approximations. We will always consider the realistic case
$$
\nu \approx 1+ \delta, \quad {\rm with}\quad \delta=4G\mu <<
1\;.\eqno(gut)$$
Since $\delta$ is of order the mass per unit length in Planck units,
it is reasonable to assume it is small (for instance it is of order
$10^{-6}$ for GUT cosmic strings).

We first consider the low energy case, just above the threshold for
pair production, with $ (k_{\perp}-2M)/2M << 1$.
In this limit  only the values of $s\approx 1$ contribute
significantly
to the integral in \(sigma), and thus we may approximate
$\epsilon\approx 1,\quad a \approx {1\over2}(1-s)(1+t),
\quad b \approx {1\over2}(1-s)(1-t),$
and then we find
$$ \sigma_m \approx
{e^2 \delta^2 \over 8\sqrt{2}\pi^{3/2}\omega}\left({k_{\perp}\over
2M}-1
\right)^{{3\over2}+|m|}{\Gamma(|m|+1) \over \Gamma(|m|+5/2)}\quad
{\rm if}\quad  {k_{\perp}\over 2M}-1 << 1 \;.\eqno(lowe)$$

At high energy, $k_{\perp}>>M$, (and $\epsilon << 1$)
the main contribution in \(sigma) arises from small values of $s$.
In this case
$$ a \approx 1-s\left(1+ \sqrt{1-t \over 1+t} \right) , \; b
\approx 1-s\left(1+ \sqrt{1+t \over 1-t} \right)$$
and one needs to be careful in finding the asymptotic behaviour. We
get
$$ \sigma_m \approx
{{e^{2}\delta^2}\over{2\pi^{2}\omega}} \int_{0}^{1}{dt}
\int_{\epsilon}^{1}{ds} {(1-t^2)^{3\over2} (1+\sqrt{1-t^2})^{-3}
\over{s^2 \sqrt{s^2-
\epsilon^2}[s^2(1+\sqrt{1-t^2})^2+\pi^2\delta^2(1-t^2)]}}\;.
\eqno(sigma2)$$
We still need to distinguish two different regimes at high energy.
The term between
square brackets in the denominator of \(sigma2) plays different
roles whether $s$ can be smaller than $\delta$ or not.
If the photon energy is not too high, $M<<k_{\perp}
<<M/\delta$, we can neglect the term proportional to
$\delta^2$, and then we find
$$ \sigma_m \approx {e^2 \delta^2 \over {1890 \pi^{2}\omega}}\left
({k_{\perp}\over M}\right )^{4}\quad {\rm if}\quad M<<k_{\perp}
<<{M\over \delta}\;.\eqno(he)$$
Finally, at  ultrahigh energies, when
$k_{\perp}>> {M\over \delta}$, the term proportional to
$\delta^2$ dominates and then we obtain
$$ \sigma_m \approx {e^2 \over {60
\pi^{4}\omega}}\left ({k_\perp\over M}\right)^2
\quad {\rm if}\quad k_{\perp}>> {M\over \delta}\;.\eqno(uhe)$$

The cross section increases with increasing photon energy. More
precisely, it increases with  increasing $k_\perp$, since
the dynamics in the direction parallel to the string is trivial.
This rather peculiar energy dependence  cannot continue unbounded,
since at sufficiently large energies it would conflict with
unitarity, which imposes a bound of order $\sigma_m\lsim \omega^{-1}$
on the partial cross sections. The unitarity limit thus restricts
the validity of eq. \(he) up to energies $k_\perp \lsim 10
M/\sqrt{e\delta}$, and that of eq. \(uhe) up to $k_\perp\lsim
100M/e$.
We think this is a consequence of the breakdown of
the validity of perturbation theory at these high energies.
Notice that for realistic values of $e$ and $\delta$, equation \(uhe)
for the ultrahigh energy behaviour of the cross section is always
in confict with unitarity.

Notice that in both high energy regimes the cross section does not
depend on the photon quantum number $m$ that determines its angular
momentum projection along the string axis. This is actually only true
up to some large value of $m$, of order $k_\perp/M$, after which the
approximations we were making break down, and the cross section
decreases with $m$. An heuristic, semiclassical explanation for this
property may arise from the following observation. The classical
analog of the cylindrical modes with given $m$ may be thought of as a
flux of particles incident upon the string from all directions with
radius of closest approach of order $\rho_{\rm min}\approx
m/k_\perp$.
\refto{Audretsch91} This will be shorter than the Compton
wavelength $1/M$ of the pair if $m<k_\perp/M$. In that case virtual
members of a pair that move along opposite sides of the string, and
thus are able to extract momentum from it, will get hit by the
photon. All values of $m$ smaller than $k_\perp/M$ should then be
equally
efficient at producing pairs, while larger values should be less
efficient.

\subhead{4.3. Cross section for ``plane wave'' photon states}

Equations \(lowe), \(he) and \(uhe) give us the cross section for
pair production in the gravitational field of a cosmic string
by a single photon in an eigenstate of the angular momentum
operator $L_3$ in eq.\(lap), with eigenvalue $l_3=\nu m\pm 1=j_3\pm
1$.
These cylindrical normal modes we  used proved to be rather
convenient
to carry out our calculations. However, it is more common and useful
to express the cross section for incoming states of definite linear
momentum, i.e. for plane waves. Of course, there are no exact plane
wave solutions in this case, since the  non-conservation of linear
momentum in the plane perpendicular to the string is precisely the
origin of the effect under  consideration. It is possible,
nevertheless,
to define  ``almost'' plane wave states, states that asymptotically
represent the superposition of an incoming plane wave distorted
by the scattered cylindrical wave. These are the
states that should be used in place of standard plane waves.
They were analysed in the context of classical and quantum scattering
problems in conical 2+1 dimensional space, both for
scalar\refto{Deser88} as well as for Dirac fields.\refto{Gerbert89a}

It is not difficult to obtain the cross section for these ``almost
plane wave'' states from the partial cross sections for states with
given $j_3=\nu m$, that we already evaluated. We have
shown how to do this in the case of scalar fields in Ref.\cite{HS91}.
Since its extension to the QED case is rather straightforward, at
least
if one is interested in the total cross section only, and not on its
polarization dependence,
let us simply outline the method developed for the scalar case.
It is based on the fact that a normal mode with the asymptotic
behaviour
$$
\psi(x)\approx  {\rm exp}(ik_\perp\rho\cos\nu\theta)
+f(\theta){{\rm exp}(ik_\perp\rho)\over\sqrt\rho}\;,\eqno(asym)$$
that represents an incoming plane wave distorted by the scattered
outgoing cylindrical wave, may be expanded in partial waves of
cylindrical type as
$$\psi(x)= \sum_{m}c_mJ_{|m|\nu}(k_\perp\rho){\rm exp}
(im\nu\theta)\;,\eqno(sum)$$
where $c_m={\rm exp}(i\pi m-i\pi|m|\nu/2)$.  Eq.\(sum) with
$\nu=1$ is of course just the usual expansion of a plane
wave in cylindrical modes. When $\nu$ is larger than
one it incorporates the scattered wave, unavoidable as the
``plane wave'' propagates in the conical topology.

Using normal modes of this sort to characterize the quantum states,
one can check that in the low energy regime the cross section for
a plane wave incoming state is dominated by the s-partial
wave, i.e. the mode with $m=0$ in \(lowe).
At high energies, on the contrary, all  partial waves contribute with
equal weight, so that $\sigma=\sum_{m}\sigma_m$, where we denote
by $\sigma$ the total cross section for a plane wave state.
This leads to a rather large enhancement of the cross section above
the value of eqs.\(he) and \(uhe), since the contribution of each
partial wave is basically independent on $m$ over a large range. Now
we must work with a better degree of approximation than in the
previous
section, to find the behaviour for very large $m$ also. It is easier
to perform the calculations doing the sum over $m$ before the
integrals on $t$ and $s$ in eq.\(sigma). One then finds that the
integrand has one extra inverse power of $s$ than before the
summation,
and that leads to a cross section at high energies for plane wave
incoming
states that is a factor of order $k_\perp/M$ larger than eqs. \(he)
and
\(uhe). We get:
$$\sigma =
\sum_{m} \sigma_{m} \approx {e^2 \delta^2\over 6720(2\pi) \omega}
\left({k_{\perp}\over M}\right )^5
\quad {\rm if}\quad M<<k_\perp<<{M\over\delta}\;.\eqno(hepw)$$
and, for the ultrahigh energy regime:
$$ \sigma =
\sum_{m} \sigma_{m} \approx {e^2 \over 60(2\pi)^3 \omega}
\left({k_{\perp}\over M}\right )^3
\quad {\rm if}\quad k_\perp>>{M\over\delta}\;.\eqno(uhepw)$$

As mentioned in the previous section, these results are in conflict
with the unitarity bound on the cross section if
$k_\perp\gsim 10M/\sqrt{e\delta}$ for eq. \(hepw) and if
$k_\perp \gsim 100M/e$ for eq. \(uhepw),
reflecting the breakdown of the perturbative expansion upon which
this derivation is based.

\head{5. CONCLUSIONS}

We have shown that the lack of global linear momentum conservation in
the plane perpendicular to a cosmic string, which is a consequence of
its conical topology, permits electron-positron pairs to be produced
by a single photon notwithstanding there is no local gravitational
field. Expressions \(lowe), \(he), \(uhe), \(hepw) and \(uhepw)
contain our quantitative results, with the cross section per unit
length
of string for this process at different energy regimes and for
alternative
incoming quantum states of the photon.

Previous results of a similar nature were already known for a
simplified model based on scalar fields.\refto{HS90,HS91,Audretsch91}
The extension to QED, though, proved not so straightforward, both in
its technical details as well as in the energy dependence
of the resulting cross sections.
The scalar model was based on a Lagrangian $\lambda\varphi\psi^2$,
with $\lambda$ a coupling constant with mass units, $\varphi$ a
massless and $\psi$ a massive scalar field. The QED result
corresponds to the scalar case with $\lambda$
replaced by $e k_\perp$. In the high energy regime this makes a
significant
difference.

The pair production process by a high energy photon in the space-time
of a cosmic string can be regarded as a consequence of a kind of
gravitational analog of the Aharonov-Bohm effect. There is also a
strictly speaking Aharonov-Bohm interaction of fermions with the
gauge
potential around a cosmic string, in models where the string carry
non-integer fluxes.\refto{Alford89a,Gerbert89b,Perkins91b}
There are many similarities in the mathematical treatment of the
Aharonov-Bohm interaction and of the gravitational effect discussed
in
this paper, but also substantial differences. Besides, the
gravitational
effect is quite insensitive to  the details of the field theory
behind
the string.

It is very striking that the cross section for pair production at
high
energies, eq.\(uhepw), grows with the energy of the incoming photon.
This behaviour should be cut-off at energies that probe the core of
the string, $k_\perp\approx \sqrt\mu$, where the metric is not well
approximated by that of eq.\(m), which has a conical singularity at
the origin. A real cosmic string has a smooth core, and its
metric approaches a flat, regular metric at the origin.
The metric is that of a snub-nosed cone.\refto{Gregory87,Garfinkle85}
It was shown, for instance, that the $1/\rho$ self-force that a
charged particle experiences in the conical space-time around a
string,\refto{Linet86,Smith90} is cut-off at a distance of order the
core radius in the snub-nosed cone metric.\refto{Perkins91a}
However, the unitarity bound on the cross section is already
violated at energies much lower than those which probe the string
core,
reflecting the breakdown at high energies of the perturbative
approach
used to derive our results for the cross section.

\head{Acknowledgments}

D.H. is grateful to the Lebedev Physical Institute and especially to
V. Frolov and the other members of the Relativity group for their
hospitality. V.S. thanks M. Castagnino and the University of Buenos
Aires as well as J.Audretsch, A.Economou and the other members of the
Relativity group at the University of Konstanz for hospitality,
collaboration and many fruitful discussions.

This work was supported by the Deutsche Forschungsgemeinschaft, the
European Community DG XII, CONICET and Fundaci\'on Antorchas.


\references
\doublespace
\refis{Aharonov59}Y. Aharonov and D. Bohm, \journal Phys. Rev.,
119,485,1959.\par
\refis{Bethe34}H. Bethe and W. Heitler, \journal Proc. Roy. Soc.,
A 146,83,1934.\par
\refis{HS90}D.D. Harari and V.D. Skarzhinsky, \plb 240,322,1990.\par
\refis{HS91}D.D. Harari and V.D. Skarzhinsky, in {\it ``Proceedings
of
the V Seminar on Quantum Gravity''}, Moscow 1990, ed. by M.A.
Markov, V.A. Berezin and V.P. Frolov, World Scientific (1991); and
in {\it ``Proceedings of
First International A. D. Sakharov Conference in Physics''}, Moscow
(1991).\par
\refis{Vilenkin85}A. Vilenkin, \journal Phys.
Rep.,121,263,1985.  \par
\refis{Gott85} J. Gott, \journal Ap. J.,288,422,1985.\par
\refis{Deser84}S. Deser, R. Jackiw and G. 't Hooft, \journal Ann.
Phys., 152,220,1984.\par
\refis{Deser88} S. Deser and R. Jackiw, \journal Comm. Math.
Phys.,118,495,1988.; G.'t Hooft, \journal Comm. Math.
Phys.,117,685,1988.\par
\refis{Henneaux84} M. Henneaux, \prd 29,2766,1984.; S. Deser, \cqg
2,489,1985.\par
\refis{Audretsch91}J. Audretsch and A. Economou, \prd 44,980,1991.
and \prd 44,3774,1991.; J.Audretsch, A. Economou and D. Tsoubelis,
\prd
45,1103,1992.\par
\refis{Voropaev91}S.A. Voropaev, D.V. Galtsov and D.A. Spasov, \plb
267,91,1991.;  \journal Europhys. Lett.,12,609,1990.\par
\refis{Aliev89}A.N. Aliev and D.V. Gal'tsov, \journal Annals of
Physics,193,142,1989.\par
\refis{Gerbert89a}Ph. de Sousa Gerbert and R. Jackiw, \journal Comm.
Math.Phys.,124,229,1989.\par
\refis{Gerbert89b}Ph. de Sousa Gerbert, \prd 40,1346,1989.\par
\refis{Weinberg72}S. Weinberg,
{\sl Gravitation and Cosmology}, Wiley, N.Y. (1972)\par
\refis{Alford89a}M.G. Alford and F. Wilczeck, \prl 62,1071,1989.\par
\refis{Alford89b}M.G. Alford, J. March-Russell and F. Wilczeck,
\npb 328,140,1989.\par
\refis{Perkins91b}W.B. Perkins, L. Perivolaropoulos, A.C. Davis, R.H.
Brandenberger and A. Matheson, \npb 353,237,1991.\par
\refis{Perkins91a}W.B. Perkins and A.C. Davis,
\npb 349,207,1991.\par
\refis{Kay91}B.S. Kay and U.M. Studer, \journal Commun. Math.
Phys.,139,103,1991.\par
\refis{Hagen90}C.R. Hagen, \prl 64,503,1990. and \prd
41,2015,1990.\par
\refis{Gradshteyn80}I.S. Gradshteyn and I.M. Ryzhik, {\it Table of
integrals, series and products}, Academic Press (1980).\par
\refis{Linet86} B. Linet, \prd 33,1833,1986.\par
\refis{Smith90} G. Smith, in {\it `` Formation and evolution of
cosmic
strings''}, Proceedings of the Cambridge Workshop, ed. by G.W.
Gibbons, S.W. Hawking and T. Vachaspati, Cambridge Univ. Press
(1990).\par
\refis{Serebryany89} E.M. Serebryany and V.D. Skarzhinsky,
``The electromagnetic radiation at the Aha\-ro\-nov\--Bohm
scat\-te\-ring" in
{\it Pro\-cee\-dings of the Lebedev Physical Institute} {\bf 197}
(1989) 181.\par
\refis{Frolov88} V.P. Frolov, E.M. Serebryany and V.D. Skarzhinsky,
in
{\it Proceedings of the 4th Moscow Seminar on Quantum Gravity},  ed.
by
M.A. Markov, V.A. Berezin and V.P. Frolov, World Sci. (1988).\par
\refis{Vilenkin81}A. Vilenkin, \prd 23,852,1981.\par
\refis{Ford81}L.Ford and A. Vilenkin, \journal J. Phys.
A,14,2353,1981.\par
\refis{Garfinkle85}D. Garfinkle, \prd 32,1323,1985.  \par
\refis{Gregory87}R. Gregory, \prl 59,740,1987.\par
\refis{Audretsch93}J. Audretsch, U. Jasper and V. Skarzhinsky, in
preparation.\par
\endreferences
\endpaper\end